\RequirePackage{ifpdf}
\ifpdf 
\documentclass[pdftex]{sigma}
\else
\documentclass{sigma}
\fi

\long\def\symbolfootnote[#1]#2{\begingroup%
\def\thefootnote{\fnsymbol{footnote}}\footnote[#1]{#2}\endgroup}
\usepackage{amsmath}
\usepackage{amsfonts}
\usepackage{amsthm}
\usepackage[T1]{fontenc}
\usepackage{mathtools}
\newtheorem{thm}{Theorem}[section]

\theoremstyle{definition}

\theoremstyle{remark}
\newtheorem{rem}[thm]{Remark}

\newcommand*{\bfrac}[2]{\genfrac{}{}{0pt}{}{#1}{#2}}

\newcommand{\hook}{\raisebox{-0.35ex}{\makebox[0.6em][r]
{\scriptsize $-$}}\hspace{-0.15em}\raisebox{0.25ex}
{\makebox[0.4em][l]{\tiny $|$}}}

\begin{document}

\renewcommand{\PaperNumber}{***}

\FirstPageHeading

\ShortArticleName{Hidden symmetries of Euclideanised Kerr-NUT-(A)dS
metrics}

\ArticleName{Hidden symmetries of Euclideanised Kerr-NUT-(A)dS metrics
in certain scaling limits}

\Author{Mihai VISINESCU~$^\dag$ and
Gabriel Eduard V\^{I}LCU~$^{\ddag,\S}$}

\AuthorNameForHeading{M. Visinescu and G.E. V\^{\i}lcu}

\Address{$^\dag$~National Institute for Physics and Nuclear
        Engineering, Department of Theoretical Physics, P.O.Box M.G.-6,
        Magurele, Bucharest, Romania}
\EmailD{mvisin@theory.nipne.ro} 


\Address{$^\ddag$~Petroleum-Gas University of Ploie\c sti,
Department of Mathematical Economics,
Bulevardul Bucure\c sti, Nr. 39, Ploie\c sti 100680, Romania}
\EmailD{gvilcu@upg-ploiesti.ro}
\Address{$^\S$~University of Bucharest,
Faculty of Mathematics and Computer Science,
Research Center in Geometry, Topology and Algebra,
Str. Academiei, Nr. 14, Sector 1, Bucharest 70109, Romania}
\EmailD{gvilcu@gta.math.unibuc.ro}

\ArticleDates{Received ???, in final form ????; Published online ????}

\Abstract{
The hidden symmetries of higher dimensional Kerr-NUT-(A)dS metrics
are investigated. In  certain scaling limits these metrics are related
to the Einstein-Sasaki ones. The complete set of Killing-Yano tensors of
the Einstein-Sasaki spaces are presented. For this purpose the
Killing forms of the Calabi-Yau cone over the Einstein-Sasaki manifold
are constructed. Two new Killing forms on  Einstein-Sasaki manifolds
are identified associated with the complex volume form of the cone
manifolds. Finally the Killing forms on mixed 3-Sasaki manifolds are
briefly described.}


\Classification{53C15; 53C25; 81T20} 


\section{Introduction}

The usual spacetimes symmetries are represented by isometries connected
with the Killing vector fields. Slightly more generally, the conformal
Killing vector field preserve a given conformal class of metrics. For
each of the (conformal) Killing vector fields there exists a conserved
quantity for the (null) geodesic motions.

Besides them a spacetime may also possess hidden symmetries generated
by higher order symmetric or antisymmetric tensor fields. The symmetric
St\" ackel-Killing tensors give rise to conserved quantities of higher
order in particle momenta. A natural generalization of (conformal)
Killing vector fields is given by the antisymmetric (conformal)
Killing-Yano tensors. Killing-Yano tensors are also called Yano tensors
or Killing forms, and conformal Killing-Yano tensors are sometimes
referred as conformal Yano tensors, conformal Killing forms or twistor
forms.

In physics, Yano tensors play a fundamental role being related to the
separability of field equations with spin, pseudoclassical spinning
models, the existence of quantum symmetry operators, supersymmetries,
etc.

In this paper we want to take a closer look at the Killing forms of
Kerr-NUT-(A)dS metrics which are
related to Einstein-Sasaki metrics. An Einstein-Sasaki manifold is a
Riemannian manifold that is both Sasakian and Einstein. In the last
time Sasakian manifolds, as an odd-dimensional analog of K\" ahler
manifolds, have become of high interest. Sasakian manifolds are contact
manifolds satisfying a normality (or integrability) condition. On the
other hand the contact geometry is motivated by classical mechanics, a
contact space corresponding to the odd-dimensional extended phase space
that includes the time variable. Recently Einstein-Sasaki geometries
have been the object of much attention in connection with the
supersymmetric backgrounds relevant to the AdS/CFT correspondence.
On the other hand a lot of interest focuses on higher dimensional black
hole spacetimes \cite{ER}. The search of hidden symmetries generated by
the Killing forms in rotating black hole geometries has an important
role for describing the properties of black holes in various dimensions.

The Kerr-NUT-AdS metrics in all dimensions have been constructed in
\cite{CLP}. The general Kerr-NUT-AdS metrics have $(2n-1)$ non-trivial
parameters where the spacetime dimension is $(2n+1)$ in the
odd-dimensional case and $(2n)$ in the even dimensional case. It was
also considered the BPS, or supersymmetric, limits of these metrics.
After Euclideanisation, these limits yield in odd dimensions new
families of Einstein-Sasaki metrics, whereas the even-dimensional
metrics result in the Ricci-flat K\"ahler manifolds. An alternative
procedure was proposed in \cite{DK} generalizing the scaling limit of
Martelli and Sparks \cite{MS}.
More precisely, in a certain limit one gets an Einstein-K\" ahler
metric from an even-dimensional Kerr-NUT-(A)dS spacetime and the
Einstein-Sasaki space is constructed as a $U(1)$ bundle over this
metric. On the other hand, performing the scaling limit of the
odd-dimensional Kerr-NUT-(A)dS spacetimes one gets directly the same
Einstein-Sasaki space obtained as a $U(1)$ bundle over the
Einstein-K\" ahler metric \cite{DK}.

The Kerr-NUT-(A)dS metrics possess explicit and hidden symmetries
encoded in a series of rank two St\"ackel-Killing tensors and Killing
vectors \cite{CLP}. These symmetries allow one constructs a set of
quantities conserved along geodesics. Moreover they are functionally
independent and in involution and guarantee complete integrability of
the geodesic motions \cite{PKVK,KKPV,SK}.

The structure of the hidden symmetries for a Sasaki space is derived
from the characteristic Sasakian $1$-form. Killing-Yano tensors
alternate closed conformal Killing-Yano tensors as the rank increases
\cite{DK}. The corresponding hidden symmetries are purely geometrical,
irrespective of the fact whether the Einstein equations are satisfied
or not.

One of the purposes of this paper is to point out the special case of
the higher dimensional Kerr-NUT-(A)dS metrics which are related to the
Einstein-Sasaki ones. In this case there are two additional
Killing-Yano tensors taking into account that the metric cone is
Calabi-Yau \cite{US}. These two exceptional Killing forms can be also
described using the Killing spinors of an Einstein-Sasaki manifold
\cite{CB}.

In the main body of the present paper we consider the Killing forms on
Einstein-Sasaki spaces. Let us remark that versions of $M$-theory could
be formulated in spacetimes with various number of time dimensions
giving rise to exotic spacetime signatures \cite{H1,H2}.
The paraquaternionic structures arise in a natural way in modern
studies in string theories, integrable systems \cite{CMMS,DW,OV}.
The counterpart in odd-dimensions of a paraquaternionic structure is
called mixed 3-structure which appear in a natural way on lightlike
hypersurfaces in paraquaternionic manifolds. For completeness we extend
the study of Killing forms on other more particular Sasaki structures.

In Section 2 we review some basic facts about the Einstein-Sasaki
spaces and their cone manifolds. In the next Section we discuss the
Killing forms on Einstein-Sasaki spaces which proceed from
Euclideanised Kerr-NUT-(A)dS metrics in certain scaling limits. We
identity two new Killing forms associated with the complex volume form
of the cone manifold. The paper ends with conclusions in Section 4. In
an appendix we briefly discuss the Killing forms on mixed 3-Sasaki
manifolds.

\section{Mathematical preliminaries}

For convenience, the mathematical concepts and results needed to study
the hidden symmetries on Einstein-Sasaki spaces are summarized in this
Section.

\subsection{Killing vector fields and their generalizations}

A vector field $X$ on a (pseudo-)Riemannian manifold $(M,g)$
is said to be a \emph{Killing vector field} if the Lie derivative of
the metric $g$ with
respect to $X$ vanishes or, equivalently, if the Levi-Civita
connection $\nabla$ of
$g$ satisfies
\[
g(\nabla_YX,Z)+g(Y,\nabla_ZX)=0\,,
\]
for all vector fields $Y$, $Z$ on $M$. A natural generalization of
Killing vector fields is given by the \emph{conformal Killing vector
fields}, i.e. vector fields with a flow preserving a given
conformal class of metrics \cite{YAN}. On the other hand, a
\emph{conformal Killing-Yano tensor} of
rank $p$ on a (pseudo-) Riemannian manifold $(M,g)$ is a $p$-form
$\omega$ which satisfies:
\begin{gather}\label{AB}
\nabla_X\omega=\frac{1}{p+1}X \hook
d\omega-\frac{1}{n-p+1}X^*\wedge d^*\omega  \,,
\end{gather}
for any vector field $X$ on $M$, where $\nabla$ is the Levi-Civita
connection of $g$, $n$ is the dimension of~$M$,~$X^*$ is the 1-form
dual to the vector field $X$ with respect to the metric~$g$,
$\hook$ is the operator dual to the wedge product and~$d^*$ is
the adjoint of the exterior derivative $d$. If $\omega$ is co-closed
in~\eqref{AB}, then we obtain the definition of a \emph{Killing-Yano
tensor} (introduced by Yano~\cite{YAN}).  It is easy to see that for
$p=1$, they are dual to Killing vector fields. Moreover, a Killing form
$\omega$ is said to be a \emph{special Killing form} if it satisfies
for some constant $c$ the additional equation
\begin{gather}\label{SKY}
\nabla_X(d\omega)=cX^* \wedge\omega \,,
\end{gather}
for any vector field $X$ on $M$.

Besides the antisymmetric generalization of the Killing vectors one
might also consider higher order symmetric tensors. A symmetric tensor
$K_{(i_1 \dots i_k)}$ obeying the equation
\[
K_{(i_1 \dots i_k;j)} = 0 \,,
\]
is called a \emph{St\"ackel-Killing tensor}. For any geodesic with 
a tangent vector $u^i$ the following object
\[
P_K = K_{i_1 \dots i_k} u^{i_1} \cdots u^{i_k} \,,
\]
is conserved.

These two generalizations of the Killing vectors could be related.
Given two Killing-Yano tensors $\omega^{i_1, \dots, i_r}$ and
$\sigma^{i_1, \dots, i_r}$ it is possible to associate with them a
St\"ackel-Killing tensor of rank $2$:
\begin{equation}\label{KYY}
K^{(\omega,\sigma)}_{ij} = \omega_{i i_2 \dots i_r}
\sigma_{j}^{~i_2 \dots i_r}+ \sigma_{i i_2 \dots i_r}
\omega_{j}^{~i_2 \dots i_r} \,.
\end{equation}
Therefore a method to generate higher order integrals of motion is to
identify the complete set of Killing-Yano tensors. The existence of 
enough integrals of motion leads to complete integrability or even
superintegrability of the mechanical system when the number of
functionally independent constants of motion is larger than its number
of degrees of freedom.
Let us mention that when a St\"ackel-Killing tensor is of the form
\eqref{KYY}, there are no quantum anomalies thanks to an integrability
condition satisfied by the Killing-Yano tensors \cite{BC}.

\subsection{Almost Hermitian manifolds and the complex volume form}

\emph{An almost (pseudo-)Hermitian structure} on a smooth manifold $M$
is a pair $(g,J)$, where $g$ is a (pseudo-)Riemannian metric on $M$
and $J$ is an almost complex structure on $M$, which is compatible
with $g$, i.e.
\[
g(JX,JY)=g(X,Y)\,,
\]
for all vector fields $X,Y$ on $M$. In this case, the triple $(M,J,g)$
is called an \emph{almost (pseudo-) Hermitian manifold}.
Moreover, if $J$ is parallel with respect to the Levi-Civita
connection $\nabla$ of $g$,
then $(M,J,g)$ is said to be a \emph{K\"{a}hler manifold (with
indefinite metric)}. We remark that on a K\"{a}hler manifold, the
associated K\"{a}hler form, i.e the alternating 2-form $\Omega$ defined by
\[
\Omega(X,Y)=g(JX,Y)\,,
\]
is closed. In local holomorphic coordinates $(z^1,...,z^m)$, the
associated K\"{a}hler form $\Omega$ can be written as
\[
\Omega=i g_{j\bar{k}}dz^j\wedge d\bar{z}^k=\sum X^*_j\wedge Y^*_j=
\frac{i}{2}\sum Z_j^*\wedge \bar{Z}_j^*\,,
\]
where $(X_1,Y_1,...,X_m,Y_m)$ is an adapted
local orthonormal field (i.e. such that $Y_j=JX_j$), and
$(Z_j,\bar{Z}_j)$ is the associated complex frame given by
\[
Z_j=\frac{1}{2}(X_j-iY_j),\,\, \bar{Z}_j=\frac{1}{2}(X_j+iY_j)\,.
\]

We also note that the dimension of an almost (pseudo-)Hermitian
manifold is necessarily even (see e.g. \cite{KN}) and, in the case of
a K\"{a}hler
manifold, there is an intimate connection between its K\"{a}hler form
and the volume form (which is just the Riemannian volume form determined by the metric) as follows
\[
d\mathcal{V}=\frac{1}{m!}\Omega^m\,,
\]
where $d\mathcal{V}$ denotes the volume form of $M$, $\Omega^m$ is the 
wedge product of $\Omega$ with itself $m$ times, $m$ being the complex
dimension of $M$ (see \cite{BAL}). Hence the volume form is a real 
$(m,m)$-form on $M$.

On the other hand, if the volume of a K\"{a}hler manifold is written as
\[
d\mathcal{V} = dV \wedge d{\bar{V}}
\]
then $dV$ is the complex volume form of $M$. It is now clear that the 
complex volume form of a K\"{a}hler manifold can be written in a 
simple way with respect to any (pseudo-)orthonormal basis, using 
complex vierbeins $e_i+Je_i$. In fact, the complex volume form of 
a K\"{a}hler manifold $M$ is, up to a power factor of the imaginary 
unit $i$,  the exterior product of these complex vierbeins.

\subsection{The K\"{a}hler cone of an Einstein-Sasaki manifold}

Let $M$ be a smooth manifold equipped with a triple
$(\varphi,\xi,\eta)$, where $\varphi$ is a field  of endomorphisms
of the tangent spaces, $\xi$ is a vector field and $\eta$ is a
1-form on $M$. If we have:
\[
\varphi^2=\tau(-I+\eta\otimes\xi),\ \ \  \eta(\xi)=1 \,,
\]
then we say that:

$(i)$ $(\varphi,\xi,\eta)$ is an \emph{almost contact
structure} on $M$, if $\tau=1$ (cf. \cite{SAS}).

$(ii)$ $(\varphi,\xi,\eta)$ is an \emph{almost paracontact
structure} on $M$, if $\tau=-1$ (cf. \cite{Sat}).

A (pseudo-)Riemannian metric $g$ on $M$ is said to be \emph{compatible}
with the
almost (para)contact structure  $(\varphi,\xi,\eta)$ if and only if
the relation
\[
g(\varphi X, \varphi Y)=\tau
[g(X,Y)-\varepsilon\eta(X)\eta(Y)] \,,
\]
holds for all pair of vector fields $X,Y$ on $M$, where
$\varepsilon = \pm1$, according as $\xi$ is space-like or time-like,
respectively.

An almost (para)contact metric structure $(\varphi,\xi,\eta,g)$ is a
\emph{(para-)Sasakian structure} if and only if the Levi-Civita
connection $\nabla$ of the metric $g$ satisfies
\begin{equation}\label{SS}
(\nabla_X\varphi)
Y=\tau[g(X,Y)\xi-\epsilon\eta(Y)X]\,,
\end{equation}
for all vector fields $X,Y$ on $M$ (see \cite{BLR}).

A (para-)Sasakian structure may also be reinterpreted and characterized
in terms of the metric cone as follows.
The (space-like) metric cone of a (pseudo-)Riemannian manifold
$(M,g)$ is the (pseudo-)Riemannian manifold $C(M)=(0,\infty)\times M$
with the metric given by
\[
\bar{g}=dr^2+r^2g\,,
\]
where $r$ is a
coordinate on $(0,\infty)$. Then $M$ is a Sasaki manifold if and only
if its metric cone $C(M)$ is  K\"{a}hler   \cite{BG} and we have a
similar characterization for para-Sasakian manifolds \cite{ACG}. In
particular, the cone $C(M)$ is equipped with an integrable
complex structure $J$ and a K\"{a}hler 2-form $\Omega$, both of which
are parallel
with respect to the Levi-Civita connection $\bar{\nabla}$ of $\bar{g}$.
Moreover, $M$ has odd dimension $2n + 1$, where $n+1$ is the complex
dimension of the K\"{a}hler cone. We note that the Sasakian manifold
$(M,g)$ is naturally isometrically embedded into the cone via the
inclusion
\[
M=\{r=1\}=\{1\}\times M\subset C(M) \,,
\]
and the K\"{a}hler structure of the cone $(C(M),\bar{g})$ induces an
almost contact metric structure $(\phi,\xi,\eta,g)$ on $M$ satisfying
\eqref{SS}.

A \emph{(para-)Einstein-Sasaki manifold} is a Riemannian manifold
$(M,g)$ that is both
(para-)\-Sasaki and Einstein, i.e. a (para-)Sasakian manifold satisfying
the Einstein condition
\[
Ric_{g} = \lambda g \,,
\]
for some real constant $\lambda$, where $Ric_{g}$ denotes the Ricci
tensor of $g$. Einstein manifolds with $\lambda=0$ are called
\emph{Ricci-flat manifolds}. Similarly, an \emph{Einstein-K\"{a}hler
manifold} is a Riemannian manifold $(M,g)$ that is both K\"{a}hler
and Einstein. The most important subclass  of Einstein-K\"{a}hler
manifolds are the \emph{Calabi-Yau manifolds}, which are K\"{a}hler
and Ricci-flat.

It is also very important to note that the Gauss equation relating
the curvature of submanifolds
to the second fundamental form shows that a Sasaki manifold $M$ is
Einstein if and only
if the cone metric $C(M)$ is  K\"{a}hler Ricci-flat. In particular the
K\"{a}hler cone
of an Einstein-Sasaki manifold has trivial canonical bundle and the
restricted holonomy group of the cone  is contained in $SU(m)$, where
$m$  denotes the complex dimension of the K\"{a}hler cone \cite{BG2,SP}.

\section{Killing forms on Einstein-Sasaki spaces}

\subsection{Progression from Einstein-K\" ahler to Einstein-Sasaki to
Calabi-Yau manifolds}

Suppose we have an Einstein-Sasaki metric $g_{ES}$ on a manifold
$M_{2n+1}$ of odd dimension $2n+1$. An Einstein-Sasaki manifold can
always be written as a fibration over an Einstein-K\"{a}hler manifold
$M_{2n}$ with the metric $g_{EK}$ twisted by the overall $U(1)$ part of
the connection \cite{GHP}
\begin{equation}\label{ES}
ds^2_{ES} = (d\psi_n + 2A)^2 + ds^2_{EK}\,,
\end{equation}
where $dA$ is given as the K\"ahler form of the Einstein-K\"ahler base.
This can be easily seen when we write the metric of the cone manifold
$M_{2n+2}=C(M_{2n+1})$ as
\[
ds^2_{cone} = dr^2 + r^2 ds^2_{ES} = dr^2 + r^2 \left((d\psi_n + 2A)^2
+ ds^2_{EK}\right)\,.
\]
The cone manifold is Calabi-Yau ( i. e.  Ricci flat and K\"ahler) and
its K\"ahler form can be written as
\[
\Omega_{cone} = r dr\wedge (d\psi_n + 2A) + r^2 \Omega_{EK}\,,
\]
and the K\"ahler condition $d\Omega_{cone} = 0 $ implies
\[
dA =  \Omega_{EK}\,.
\]

The Sasakian 1-form of the Einstein-Sasaki metric is
\[
\eta=2A +d\psi_n \,,
\]
which is a special unit-norm Killing 1-form obeying
for all vector fields $X$ \cite{US}
\begin{eqnarray}
\nabla_X \eta &=& \frac{1}{2} X \hook d\eta \,,\nonumber\\
\nabla_X (d\eta) &=& -2 X^* \wedge \eta \,.\nonumber
\end{eqnarray}

\subsection{Kerr-NUT-(A)-dS space in a certain scaling limit}

In recent time new Einstein-Sasaki spaces have been constructed by
taking certain BPS \cite{CLPP} or scaling limits \cite{MS,DK} of the
Euclideanised Kerr-de Sitter metrics.

In even dimensions, performing the scaling limit on the Euclideanised
Kerr-NUT-(A)dS spaces, the Einstein-K\"ahler metric
$g_{EK}$ and  the K\"ahler potential $A$ are \cite{DK}:
\begin{eqnarray}
g_{EK}&=&\frac{\Delta_\mu dx_{\mu}^2}{X_\mu}
+\frac{X_\mu}{\Delta_\mu}\left (\sum_{j=0}^{n-1}
\sigma_{\mu}^{(j)} d\psi_j\,\right )^{2}\,,\nonumber\\
X_\mu&=&-4\prod_{i=1}^{n+1}(\alpha_i-x_\mu)-2b_\mu\,,\nonumber\\
A &=&\sum_{k=0}^{n-1} \sigma^{(k+1)} d\psi_k\,, \nonumber
\end{eqnarray}
with

\begin{gather}
\Delta_{\mu}=\prod_{\nu\ne\mu}(x_{\nu}-x_{\mu})\,,\nonumber\\
\sigma^{(k)}_\mu=\!\!\!\!\!\sum_{\substack{\nu_1<\dots<\nu_k
\\\nu_i\ne\mu}}\!\!\!\!\!x_{\nu_1}\dots x_{\nu_k},\quad
\sigma^{(k)}=\!\!\!\!\!\sum_{\nu_1<\dots<\nu_k}\!\!\!\!\!x_{\nu_1}\dots
x_{\nu_k}\,.\nonumber
\end{gather}

Here, coordinates $x_{\mu} \, (\mu = 1, \dots , n)$ stands for the Wick
rotated radial coordinate and longitudinal angles and the Killing
coordinates $\psi_k \, (k= 0, \dots, n-1)$ denote time and azimuthal
angles with Killing vectors $\xi^{(k)} = \partial_{\psi_k}$. Also
$\alpha_i \, (i = 1, \dots, n+1)$ and $b_{\mu}$ are constants related
to the cosmological constant, angular momenta, mass and NUT parameters
\cite{CLP}.

We mention that in the case of odd-dimensional Kerr-NUT-(A)dS spaces
the appropriate scaling limit leads to the same Einstein-Sasaki metric
\eqref{ES}.

The hidden symmetries of the Sasaki manifold $M_{2n+1}$ are described by
the special Killing $(2k+1)-$forms \cite{US}:
\begin{equation}\label{kfes}
\Psi_k = \eta \wedge (d\eta)^k \quad , \quad k = 0, 1,
\cdots , n-1\,.
\end{equation}
A sketch of this assertion is given in the appendix in a more general
context.

Semmelmann obtained in \cite{US} that special Killing forms
on a Riemannian manifold $M$ are exactly those forms which translate
into parallel forms on the metric cone $C(M)$. Therefore, the metric
cone being
either flat or irreducible, the problem of finding all special Killing
forms is reduced to a holonomy
problem (see \cite{BES}). In the case of holonomy $U(n+1)$, i.e. the
cone $M_{2n+2}=C(M_{2n+1})$ is K\"{a}hler, or equivalently $M_{2n+1}$
is Sasaki, it follows that all special Killing forms are
spanned by the forms $\Psi_k$ defined above.
Besides these Killing forms, there are $n$ closed conformal Killing
forms (also called $\ast$-Killing forms)
\[
\Phi_k = (d\eta)^k \quad , \quad k = 1, \cdots , n \,.
\]

Moreover, in the case of holonomy $SU(n+1)$, i.e. the cone
$M_{2n+2}=C(M_{2n+1})$ is K\"{a}hler and Ricci-flat, or equivalently
$M_{2n+1}$ is Einstein-Sasaki, it follows that we have
{\it two additional} Killing forms of degree $n+1$ on the manifold
$M_{2n+1}$. These additional
Killing forms are connected with the additional parallel forms of the
Calabi-Yau cone manifold $M_{2n+2}$ given by the complex volume form
and its conjugate \cite{US}.

In order to write explicitly these additional Killing forms, we
introduce the complex vierbeins on the Einstein-K\"ahler manifold
$M_{2n}$. First of all we shall write the metric $g_{EK}$ in the form
\[
g_{EK}= o^{\hat \mu} o^{\hat \mu}+
{\tilde o}^{\hat \mu}{\tilde o}^{\hat \mu}\,,
\]
and the K\"ahler 2-form
\[
\Omega=dA=o^{\hat \mu}\wedge {\tilde o}^{\hat \mu}\,.
\]
where
\begin{gather}
o^{\hat \mu} =\sqrt{\frac{\Delta_\mu}{X_\mu(x_\mu)}}\, dx_\mu
\,,\nonumber\\
{\tilde o}^{\hat \mu} = \sqrt{\frac{X_\mu(x_\mu)}{\Delta_\mu}}
 \sum_{j=0}^{n-1}\sigma_{\mu}^{(j)} d\psi_j\,.\nonumber
\end{gather}

We introduce the following complex vierbeins on Einstein-K\"ahler
manifold $M_{2n}$:
\[
\zeta_\mu = o^{\hat \mu} + i {\tilde o}^{\hat \mu}\quad , \quad \mu = 1,
\cdots , n\,.
\]

On the Calabi-Yau cone manifold $M_{2n+2}$ we take
$\Lambda_{\mu} = r \zeta_\mu$ for $\mu = 1, \cdots , n$ and
\[
\Lambda_{n+1} = \frac{dr}{r} + i \eta\,.
\]

The standard complex volume form of the Calabi-Yau cone manifold
\cite{YO} $M_{2n+2}$ is
\[
dV = \Lambda_1 \wedge \Lambda_2 \wedge \cdots \wedge \Lambda_{n+1}\,.
\]

As real forms we obtain the real respectively the imaginary part
of the complex volume form. For example, writing
\[
\Lambda_j = \lambda_{2j-1} +i \lambda_{2j},\, j=1,...,n+1\,,
\]
we obtain that the real part of the complex volume is given by
\begin{equation}\label{REV}
\mathcal{R}e\,\, dV =\sum_{p=0}^{[\frac{n+1}{2}]}
\sum_{\bfrac{1\leq i_1<i_2<...<i_{n+1}\leq 2n+2}{(C)}}(-1)^p 
\lambda_{i_1}
\wedge \lambda_{i_2}\wedge...\wedge \lambda_{i_{n+1}} \,,
\end{equation}
where the condition $(C)$ in \eqref{REV} means that in the second
sum are taken
only the indices $ i_1,...,i_{n+1}$ such that $i_1+...+i_{n+1}=
(n+1)^2+2p$ and $(i_k,i_{k+1})\neq(2j-1,2j)$, for all
$k\in\{1,...,n\}$ and $j\in\{1,...,n+1\}$.

On the other hand, we obtain that the imaginary part of the complex
volume is given by
\begin{equation}\label{IMV}
\mathcal{I}m\,\, dV =\sum_{p=0}^{[\frac{n}{2}]}
\sum_{\bfrac{1\leq i_1<i_2<...<i_{n+1}\leq 2n+2}{(C')}}(-1)^p \lambda_{i_1}
\wedge \lambda_{i_2}\wedge...\wedge \lambda_{i_{n+1}} \,,
\end{equation}
where the condition $(C')$ in \eqref{IMV} means that in the second sum 
are considered
only the indices $ i_1,i_2,...,i_{n+1}$ such that $i_1+...+i_{n+1}=
(n+1)^2+2p+1$ and $(i_k,i_{k+1})\neq(2j-1,2j)$, for all
$k\in\{1,...,n\}$ and $j\in\{1,...,n+1\}$.

Finally, the Einstein-Sasaki manifold $M_{2n+1}$ is identified with the
submanifold $\{r=1\}$ of the Calabi-Yau cone manifold
$M_{2n+2}=C(M_{2n+1})$.
The additional Killing forms on the Einstein-Sasaki spaces 
are connected  with the  parallel forms on the metric cone. 
For this purpose we make use of the fact that for any $p$-form 
$\omega^M$ on the space $M_{2n+1}$ we can define an associated 
$(p+1)$-form $\omega^C$ on the cone $C(M_{2n+1})$
\[
\omega^C := r^p d r \wedge \omega^M + \frac{r^{p+1}}{p+1} d\omega^M \,.
\]
Moreover $\omega^C$ is parallel if and only if $\omega^M$ is a special 
Killing form \eqref{SKY} with constant $c= -(p+1)$ \cite{US}.
The 1-1-correspondence between special Killing $p$-forms on 
$M_{2n+1}$ and parallel $(p+1)$-forms on the metric cone $C(M_{2n+1})$
allows us to describe the additional Killing forms on Einstein-Sasaki 
spaces.

Therefore in order to find  the additional Killing forms on the 
manifold $M_{2n+1}$ we must identify the $\omega^M$
form in the complex volume form of the Calabi-Yau cone. An explicit 
example is presented in \cite{MV} in connection with the 
five-dimensional $Y(p,q)$ spaces \cite{GMSW1,GMSW2}.

\section{Conclusions}

In this paper we presented the complete set of Killing forms on
Einstein-Sasaki spaces associated with Euclideanised Kerr-NUT-(A)dS
spaces in a certain scaling limit. The multitude of Killing-Yano and
St\" ackel-Killing tensors makes possible a complete integrability of
geodesic equations.
In the case of geodesic and Klein-Gordon equations, the existence of
separable coordinates is connected with St\"ackel-Killing tensors. On
the other hand from (conformal) Killing-Yano tensors one can construct
first order differential operators which commute with Dirac operators
\cite{CML}. In \cite{OY1,CKK} it was shown that the solutions of Dirac
equation in general higher dimensional Kerr-NUT-(A)dS spacetimes can be
found by separating variables and the resulting ordinary differential
equations can be completely decoupled. It is interesting to study
separability and eigenvalues of Dirac operators on Einstein-Sasaki
manifolds.
Let us note also that in the higher dimensional Kerr-NUT-(A)dS
spacetimes the stationary string equations are completely integrable
\cite{KF}.
An important open question is a separability problem for the 
gravitational perturbations in higher dimensional rotating black holes 
spacetimes, some preliminary results being achieved recently \cite{OY2}.

Another important direction of research is whether the Killing forms 
are also intrinsically linked to other higher spin perturbations. It is 
still an open question whether massless field equations, e. g. the 
Maxwell field, allow separation of variables in Kerr-NUT-(A)dS spaces.

These remarkable properties of higher dimensional black hole solutions
offer new perspectives in investigation of hidden symmetries of other
spacetimes structures. As a possible extension of these techniques we
present in an appendix the case of spaces with mixed 3-structures which
appear in many modern studies.
Finally we mention some recent extensions of the Killing-Yano symmetry 
in the presence of skew-symmetric torsion. Preliminary results 
\cite{HKWY1,HKWY2} indicate that Killing forms in the presence of 
torsion preserve most of the properties of the standard Killing forms.

\appendix
\section{Killing forms on mixed 3-Sasakian manifolds}

The study of 3-Sasakian manifolds was initiated by Kuo \cite{KUO} and
presently there is an extensive literature on this topic
(see for example \cite{BGM} and references therein).
It is well known that these manifolds are of great interest in physics,
owing to their applications in supergravity and M-theory
\cite{ACA,AGR,GIB} and there exists a close relationship between
quaternionic K\"{a}hler and 3-Sasakian structures \cite{KO}.
On the other hand, the theory of paraquaternionic K\"{a}hler manifolds
parallels the theory of quaternionic K\"{a}hler manifolds, but it uses
the algebra of paraquaternionic numbers, in which two generators have
square 1 and one generator has square -1 \cite{GRM}. In what follows
we recall some basic facts concerning this kind of structures,
together with their closely linked counterpart in odd dimension
(mixed 3-Sasakian structures).

An \emph{almost para-hypercomplex structure} on a smooth manifold
$M$ is a triple $H=(J_1,J_2,J_3)$ of $(1,1)$-type tensor fields on
$M$ satisfying:
\[
J_\alpha^2=-\tau_\alpha {\rm Id},\ J_\alpha J_\beta=
-J_\beta J_\alpha=\tau_\gamma J_\gamma\,,
\]
for any $\alpha\in\{1,2,3\}$ and for any even permutation
$(\alpha,\beta,\gamma)$ of $(1,2,3)$, where $\tau_1=
\tau_2=-1=-\tau_3$.
In this case $(M,H)$ is said to be an \emph{almost para-hypercomplex
manifold}.
A semi-Riemannian metric $g$ on $(M,H)$ is said to be
\emph{compatible} or \emph{adapted} to the almost para-hypercomplex
structure $H=(J_{\alpha})_{\alpha=1,2,3}$ if it satisfies:
\[
g(J_\alpha X,J_\alpha Y)=\tau_{\alpha} g(X,Y)\,,
\]
for all vector fields $X$,$Y$ on $M$ and $\alpha\in\{1,2,3\}$.
Moreover, the pair $(g,H)$ is called an \emph{almost
para-hyperhermitian structure} on $M$ and the triple $(M,g,H)$ is
said to be an \emph{almost para-hyperhermitian manifold}. We note
that any almost para-hyperhermitian manifold is of dimension $4m,\
m\geq 1$, and any adapted metric is necessarily of neutral signature
$(2m,2m)$. If $\lbrace{J_1,J_2,J_3}\rbrace$ are parallel with respect
to the Levi-Civita connection of $g$, then the manifold is called
\emph{para-hyper-K\"{a}hler}.

An \emph{almost paraquaternionic Hermitian manifold} is a triple
$(M,\sigma,g)$, where $M$ is a smooth manifold, $\sigma$ is an
almost paraquaternionic structure on $M$, \emph{i.e.} a rank
3-subbundle of $End(TM)$ which is locally spanned by an almost
para-hypercomplex structure $H=(J_{\alpha})_{\alpha=1,2,3}$ and $g$
is a compatible metric with respect to $H$.  If $(M,\sigma,g)$ is an
almost paraquaternionic Hermitian manifold
such that the bundle $\sigma$ is preserved by the Levi-Civita
connection $\nabla$ of $g$, then $(M,\sigma,g)$ is said to be a
\emph{paraquaternionic K\"{a}hler manifold} \cite{GRM}. We
note that the prototype of paraquaternionic K\"{a}hler manifold is
the paraquaternionic projective space $P^n(\widetilde{\mathbb{H}})$
as described by Bla\v{z}i\'{c} \cite{BLZ}.

The counterpart in odd dimension of a
paraquaternionic structure was introduced in \cite{IMV} under the
name of \emph{mixed 3-structure}. This concept
has been refined in \cite{CP}, where the authors have introduced
positive and negative  metric mixed 3-structures. A mixed 3-structure
on a smooth manifold $M$ is a triple of
structures $(\varphi_\alpha,\xi_\alpha,\eta_\alpha)$,
$\alpha\in\{1,2,3\}$, which are almost paracontact structures for
$\alpha=1,2$ and almost contact structure for $\alpha=3$, satisfying
the following compatibility conditions
\[
\eta_\alpha(\xi_\beta)=0\,,
\]
\[
\varphi_\alpha(\xi_\beta)=\tau_\beta\xi_\gamma,\ \
\varphi_\beta(\xi_\alpha)=-\tau_\alpha\xi_\gamma \,,\\
\]
\[
\eta_\alpha\circ\varphi_\beta=-\eta_\beta\circ\varphi_\alpha=
\tau_\gamma\eta_\gamma\,,\\
\]
\[
\varphi_\alpha\varphi_\beta-\tau_\alpha\eta_\beta\otimes\xi_\alpha=
-\varphi_\beta\varphi_\alpha+\tau_\beta\eta_\alpha\otimes\xi_\beta=
\tau_\gamma\varphi_\gamma\,,
\]
where $(\alpha,\beta,\gamma)$ is an even permutation of $(1,2,3)$
and $\tau_1=\tau_2=-\tau_3=-1$.

Moreover, if a manifold $M$ with a mixed 3-structure
$(\varphi_\alpha,\xi_\alpha,\eta_\alpha)_{\alpha=\overline{1,3}}$
 admits a semi-Rie\-man\-ni\-an metric $g$ such that:
\begin{equation}\label{V7}
g(\varphi_\alpha X, \varphi_\alpha Y)=\tau_\alpha
[g(X,Y)-\varepsilon_\alpha\eta_\alpha(X)\eta_\alpha(Y)]\,,
\end{equation}
for all $X,Y\in\Gamma(TM)$ and $\alpha=1,2,3$, where
$\varepsilon_\alpha=g(\xi_\alpha,\xi_\alpha)=\pm1$, then we say that
$M$ has a \emph{metric mixed 3-structure} and $g$ is called a
\emph{compatible metric}.

In what follows a  metric mixed 3-structure will
be denoted simply with $(\varphi_\alpha,\xi_\alpha,\eta_\alpha,g)$,
leaving the condition
$\alpha\in\{1,2,3\}$ understood. We note that if
$(M,\varphi_\alpha,\xi_\alpha,\eta_\alpha,g)$
is a manifold with a metric mixed 3-structure then from \eqref{V7}
it follows
\[
g(\xi_1,\xi_1)=g(\xi_2,\xi_2)=-g(\xi_3,\xi_3)\,.
\]

Hence the vector fields $\xi_1$ and $\xi_2$ are both either
space-like or time-like and these force the causal character of the
third vector field $\xi_3$. We may therefore distinguish between
\emph{positive} and \emph{negative metric mixed 3-structures},
according as $\xi_1$ and $\xi_2$ are both space-like, or both
time-like vector fields. Because one can check that, at each point
of $M$, there always exists a pseudo-orthonormal frame field given by
$\{(E_i,\varphi_1 E_i, \varphi_2 E_i, \varphi_3
E_i)_{i=\overline{1,n}}\,, \xi_1, \xi_2, \xi_3\}$ we conclude that
the dimension of the manifold is $4n+3$ and the signature of $g$ is
$(2n+1,2n+2)$, where we put first the minus signs, if the metric
mixed 3-structure is positive (\emph{i.e.}
$\varepsilon_1=\varepsilon_2=-\varepsilon_3=1$), or the signature of
$g$ is $(2n+2,2n+1)$, if the metric mixed 3-structure is negative
(\emph{i.e.} $\varepsilon_1=\varepsilon_2=-\varepsilon_3=-1$).

A manifold $M$ endowed with a (positive/negative) metric
mixed 3-structure
$(\varphi_\alpha,\xi_\alpha,\eta_\alpha,g)$ is said to be a
\emph{(positive/negative) mixed 3-Sasakian structure} if
$(\varphi_3,\xi_3,\eta_3,g)$ is a Sasakian structure, while both
structures
$(\varphi_1,\xi_1,\eta_1,g)$ and $(\varphi_2,\xi_2,\eta_2,g)$ are
para-Sasakian, i.e.
\begin{equation}\label{AAA2}
(\nabla_X\varphi_\alpha)
Y=\tau_\alpha[g(X,Y)\xi_\alpha-\epsilon_\alpha\eta_\alpha(Y)X] \,,
\end{equation}
for all vector fields $X,Y$ on $M$ and $\alpha=1,2,3$.

It is important to note that, like their Riemannian
counterparts,
mixed 3-Sasakian structures are Einstein, but now the scalar
curvature can be either positive or negative.

\begin{thm}\label{cp}\cite{CP,IV} Any $(4n+3)-$dimensional manifold
endowed with a
mixed $3$-Sasakian structure is an Einstein space with Einstein
constant $\lambda=(4n+2)\theta$, with $\theta=\mp1$,
according as the metric mixed 3-structure is positive or negative,
respectively.
\end{thm}

We recall that the canonical
example of manifold with negative mixed 3-Sasakian structure is the
unit pseudo-sphere
$S^{4n+3}_{2n+2}\subset\mathbb{R}^{4n+4}_{2n+2}$, while
the pseudo-hyperbolic space
$H^{4n+3}_{2n+1}\subset\mathbb{R}^{4n+4}_{2n+2}$ can be endowed with
a canonical positive mixed 3-Sasakian structure.  We also note that
the existence of both positive and negative mixed
3-Sasakian structures in a principal $SO(2,1)$-bundle over a
paraquaternionic K\"{a}hler manifold has been recently proved in
\cite{VV}.

\begin{rem}
It is known \cite{IVV} that on a  mixed
3-Sasakian manifold $(M,\varphi_\alpha,\xi_\alpha,\eta_\alpha,g)$ of
dimension $(4n+3)$
there exists space-like, time-like and light-like Killing vector
fields. Moreover, $\eta_\alpha$ are conformal Killing-Yano
tensors of
rank~$1$ on $M$, while $d\eta_\alpha$ are strictly conformal
Killing-Yano
tensors of rank~$2$ on $M$, for $\alpha=1,2,3$. On the other hand,
the wedge products of $\eta_\alpha$ and $(d\eta_\alpha)^k$ provide
Killing $(2k +1)$-form, for $k = 0,1,\dots,2n+1$, since for any vector
field $X$ on  $M$  we have
\begin{eqnarray}
\nabla_X(\eta_\alpha\wedge \left(d\eta_\alpha\right)^k)
&=&\nabla_X\eta_\alpha\wedge\left(d\eta_\alpha\right)^k+
\eta_\alpha\wedge\nabla_X\left(d\eta_\alpha\right)^k\nonumber\\
&=&\frac{1}{2}\left(X\hook d\eta_\alpha\right)
\wedge\left(d\eta_\alpha\right)^k+k\eta_\alpha\wedge
\nabla_Xd\eta_\alpha\wedge\left(d\eta_\alpha\right)^{k-1}\nonumber\\
&=&\frac{1}{2(k+1)}X\hook\left(d\eta_\alpha\right)^{k+1}
-\frac{k}{4n+2}\eta_\alpha\wedge \left(X^*\wedge d^*(d\eta_\alpha)
\right)\wedge\left(d\eta_\alpha\right)^{k-1}\nonumber\\
&=&\frac{1}{2(k+1)}X\hook\left(d\eta_\alpha\right)^{k+1}\nonumber\\
&=&\frac{1}{2(k+1)}X\hook d\left(\eta_\alpha\wedge
(d\eta_\alpha)^{k}\right)\,.\nonumber
\end{eqnarray}

It follows from a simple computation that the wedge product
of $\eta_\alpha$ and $(d\eta_\alpha)^k$ provides a special Killing form,
since it satisfies the additional  equation
\[
\nabla_X(d(\eta_\alpha\wedge \left(d\eta_\alpha\right)^k))=-2(k+1)X^*
\wedge\eta_\alpha\wedge \left(d\eta_\alpha\right)^k \,,
\]
for any vector field $X$ on  $M$. Therefore, as in 3-Sasakian case
\cite{US}, we obtain that any linear combination of the forms
$\Psi_{k_1,k_2,k_3}$ defined by
\begin{eqnarray}
\Psi_{k_1,k_2,k_3}&=&\frac{k_1}{k_1+k_2+k_3} [\eta_1\wedge
\left(d\eta_1\right)^{k_1-1}]\wedge \left(d\eta_2\right)^{k_2}
\wedge\left(d\eta_3\right)^{k_3} \nonumber\\
&&+\frac{k_2}{k_1+k_2+k_3} \left(d\eta_1\right)^{k_1}\wedge
[\eta_2\wedge\left(d\eta_2\right)^{k_2-1}]\wedge
\left(d\eta_3\right)^{k_3} \label{kf3}\\
&&+\frac{k_3}{k_1+k_2+k_3} \left(d\eta_1\right)^{k_1}\wedge
\left(d\eta_2\right)^{k_2}\wedge[\eta_3\wedge
\left(d\eta_3\right)^{k_3-1}] \,,\nonumber
\end{eqnarray}
for arbitrary positive integers $k_1,k_2,k_3$, is a special
Killing form on $M$. The special Killing forms \eqref{kfes} could be
recovered as a particular case of \eqref{kf3} for two vanishing
integers $k_i$.
\end{rem}

\begin{rem}
For the rest of this section we consider that
$(\bar{M},\sigma,\bar{g})$ is an almost paraquaternionic
Hermitian manifold of dimension $(4n+4)$ and $(M,g)$ is an
orientable non-degenerate hypersurface of $M$ with $g=\bar{g}_{|M}$,
such that the normal bundle $TM^\perp$ is generated by a unit
space-like or time-like vector field $\xi$ normal to $M$. Then for
any vector field $X$ on $M$ and any local basis
$H=(J_{\alpha})_{\alpha=1,2,3}$ of $\sigma$, we have the decomposition
\[
J_\alpha X=\varphi_\alpha X+F_\alpha X\,,
\]
for $\alpha\in\{1, 2, 3\}$ , where $\varphi_\alpha X$ and
$F_\alpha X$ are the tangent part and the normal part of
$J_\alpha X$, respectively.
We can remark that, in fact,  $F_\alpha X\in\Gamma(TM^\perp) $
for any vector
field $X$ on $M$, and therefore we deduce the
decomposition
\[
J_\alpha X=\varphi_\alpha X+\eta_\alpha(X)\xi\,,
\]
where
\[
\eta_\alpha(X)=-\varepsilon\bar{g}(X,J_\alpha \xi),\,\varepsilon=
g(\xi,\xi)\,.
\]

If we define now the vector field $\xi_\alpha$ by
\[
\xi_\alpha=-\tau_\alpha J_\alpha \xi\,,
\]
for $\alpha\in\{1,2,3\}$, then we obtain by direct computations
that the paraquaternionic structure $\sigma$ on $\bar{M}$ induces a
positive/negative metric mixed 3-structure
$(\varphi_\alpha,\xi_\alpha,\eta_\alpha,g)$ on $M$ as follows
(see \cite{IV2} for the proof in the case of negative metric
mixed 3-structure).
\end{rem}
\begin{thm}
Let $(M,g)$ be an orientable non-degenerate hypersurface of an
almost paraquaternionic Hermitian manifold
$(\bar{M},\sigma,\bar{g})$ with the normal bundle $TM^\perp$ spanned
by a unit space-like or time-like normal vector field $\xi$. Then
$(\varphi_\alpha,\xi_\alpha,\eta_\alpha,g)$ defined above is a
positive/negative metric mixed 3-structure on $M$, according as the
generator $\xi$ is a time-like or a space-like vector field.
\end{thm}

We recall now that if $\bar{\nabla}$ is the Levi-Civita connection
on $\bar{M}$ and denote by $\nabla$ the Levi-Civita connection
induced on $M$,
then the Gauss and Weingarten formulas are given by:
\begin{equation}\label{G16}
\overline{\nabla}_XY=\nabla_XY+h(X,Y)\xi \,,
\end{equation}
and
\begin{equation}\label{W17}
\overline{\nabla}_X\xi=-AX \,,
\end{equation}
for all vector fields $X,Y$ tangent to $M$, where $h$ is the second
fundamental form of $M$ and $A$ is the fundamental tensor of
Weingarten with respect to the unit space-like or time-like normal
vector field $\xi$.

From \eqref{G16} and \eqref{W17} we deduce
\begin{equation}\label{L18}
\varepsilon h(X,Y)=g(AX,Y) \,
\end{equation}
for all $X,Y\in\Gamma(TM)$, where $\varepsilon=g(\xi,\xi)$.

\begin{thm}
Let $M$ be an orientable non-degenerate hypersurface of a
para-hyper-K\"{a}hler manifold
$(\bar{M},H=(J_1,J_2,J_3),\bar{g})$ and let
$(\varphi_\alpha,\xi_\alpha,\eta_\alpha,g)$ be the canonical metric
mixed 3-structure on $M$. Then:
\begin{enumerate}
\item[(i.)] $\eta_1,\eta_2,\eta_3$ are Killing if and only if
\[
h(X,\varphi_\alpha Y)=-h(\varphi_\alpha X, Y),\,
\alpha=1,2,3\,;
\]
\item[(ii.)] $\varphi_1,\varphi_2, \varphi_3$ are covariant constant,
provided that $M$ is a totally geodesic hypersurface of $\bar{M}$;
\item[(iii.)] $\varphi_\alpha$ is Killing if and only if $h$ is
proportional to $\eta_\alpha\otimes\eta_\alpha$, $\alpha=1,2,3$,
provided that $J_1, J_2, J_3$ are Killing.
\end{enumerate}
\end{thm}
\begin{proof}

Since each $J_\alpha$ is parallel with respect to the Levi-Civita
connection of $\bar{g}$, then using \eqref{G16}, \eqref{W17} and
\eqref{L18} we obtain
\begin{eqnarray}
0&=&(\bar{\nabla}_X J_\alpha)Y\nonumber\\&=&\bar{\nabla}_X
J_\alpha Y-J_\alpha\bar{\nabla}_XY \nonumber\\
&=&\bar{\nabla}_X (\varphi_\alpha Y+\eta_\alpha(Y)\xi)-
J_\alpha(\nabla_XY+h(X,Y)\xi)\nonumber\\
&=&(\nabla_X\varphi_\alpha)Y+\tau_\alpha h(X,Y)\xi_\alpha-
\eta_\alpha(Y)AX
+[h(X,\varphi_\alpha Y)+(\nabla_X\eta_\alpha)Y]\xi\,,\nonumber
\end{eqnarray}
for all vector fields $X,Y$ on $M$.
Taking now the tangential and the normal component of both sides of
the above equation we deduce
\begin{equation}\label{TC}
(\nabla_X\varphi_\alpha)Y=-\tau_\alpha h(X,Y)\xi_\alpha+
\eta_\alpha(Y)AX \,,
\end{equation}
and
\begin{equation}\label{NC}
(\nabla_X\eta_\alpha)Y=-h(X,\varphi_\alpha Y)\,.
\end{equation}

The proof of the assertions (i.), (ii.) and (iii.) follows now easily
using \eqref{TC} and \eqref{NC} with some standard algebraic
manipulations.
\end{proof}

Next we suppose that $M$ is a non-degenerate totally umbilical
hypersurface of a para-hyper-K\"{a}hler manifold
$(\bar{M},H=(J_1,J_2,J_3),\bar{g})$, i.e. for all vector fields
$X,Y$ on $M$ we have
\begin{equation}\label{UMB}
h(X,Y)=\lambda g(X,Y)\,,
\end{equation}
for some function $\lambda$. Now we are able to prove that the
canonical metric mixed 3-structure on $M$ can be a positive or a
negative mixed 3-Sasakian structure in some conditions (compare
with \cite{SS} for the corresponding result in quaternionic setting).

\begin{thm}
Let $M$ be an orientable non-degenerate hypersurface of a
para-hyper-K\"{a}hler manifold
$(\bar{M},H=(J_1,J_2,J_3),\bar{g})$ with the normal bundle
$TM^\perp$ spanned by a unit space-like or time-like vector field $\xi$.
If $M$ is a totally umbilical hypersurface of $\bar{M}$, then:
\begin{enumerate}
\item[(i.)] the canonical metric mixed 3-structure
$(\varphi_\alpha,\xi_\alpha,\eta_\alpha,g)$ on $M$ is a positive
mixed 3-Sasakian structure if and only if $\lambda=-1$ and
$\xi$ is time-like.
\item[(ii.)] the canonical metric mixed 3-structure
$(\varphi_\alpha,\xi_\alpha,\eta_\alpha,g)$ on $M$ is a negative
mixed 3-Sasakian structure if and only if $\lambda=-1$ and $\xi$
is space-like.
\end{enumerate}
\end{thm}
\begin{proof}

Using \eqref{UMB} in \eqref{TC} we obtain
\begin{equation}\label{TC2}
(\nabla_X\varphi_\alpha)Y=-\tau_\alpha \lambda g(X,Y)\xi_\alpha+
\eta_\alpha(Y)AX\,,
\end{equation}
for all vector fields $X,Y$ on $M$.

On the other hand, from \eqref{L18} and \eqref{UMB} we derive
\begin{equation}\label{TC3}
AX=\varepsilon \lambda X\,,
\end{equation}
for any vector field $X$ on $M$.

From \eqref{TC2} and \eqref{TC3} we deduce
\begin{equation}\label{TC4}
(\nabla_X\varphi_\alpha)Y=-\lambda\tau_\alpha[g(X,Y)\xi_\alpha-
\frac{\varepsilon}{\tau_\alpha}\eta_\alpha(Y)X] \,,
\end{equation}
for all vector fields $X,Y$ on $M$.

(i.) Taking now into account that in the case of a positive metric
mixed 3-structure we have $\varepsilon_\alpha\tau_\alpha=-1$, for
$\alpha=1,2,3$, we obtain that the equation \eqref{TC4} can be
rewritten as
\begin{equation}\label{TC5}
(\nabla_X\varphi_\alpha)Y=-\lambda\tau_\alpha[g(X,Y)\xi_\alpha+
\varepsilon\varepsilon_\alpha\eta_\alpha(Y)X]\,.
\end{equation}
Comparing \eqref{AAA2}  with \eqref{TC5}  we deduce that the canonical
metric mixed 3-structure $(\varphi_\alpha,\xi_\alpha,\eta_\alpha,g)$
on $M$ is a positive mixed 3-Sasakian structure if and only if
$\lambda=-1$ and $\varepsilon=-1$, and the assertion is now clear.

(ii.) Since in the case of a  negative metric  mixed 3-structure we
have $\varepsilon_\alpha\tau_\alpha=1$, for $\alpha=1,2,3$, we can
rewrite \eqref{TC4} in  the following form:
\begin{equation}\label{TC6}
(\nabla_X\varphi_\alpha)Y=-\lambda\tau_\alpha[g(X,Y)\xi_\alpha-
\varepsilon\varepsilon_\alpha\eta_\alpha(Y)X]\,.
\end{equation}
Comparing now \eqref{AAA2}  and \eqref{TC6}  we deduce that the
canonical metric mixed 3-structure $(\varphi_\alpha,\xi_\alpha,
\eta_\alpha,g)$ on $M$ is a negative mixed 3-Sasakian structure
if and only if $\lambda=-1$ and $\varepsilon=1$, and the conclusion
follows.
\end{proof}

\section*{Acknowledgments}
The authors would like to thank to the referees for comments and 
valuable suggestions. MV was supported by
CNCS-UEFISCDI, project number PN-II-ID-PCE-2011-3-0137. The work of
GEV  was supported by CNCS-UEFISCDI, project number
PN-II-ID-PCE-2011-3-0118.


\begin{thebibliography}{99}
\footnotesize\itemsep=0pt
\providecommand{\eprint}[2][]{\href{http://arxiv.org/abs/#2}{arXiv:#2}}
%
\bibitem{ACA}  Acharya B.S.,  Figueroa-O'Farrill J.M.,  Hull C. M.,
Spence B. J., Branes at conical singularities
and holography, \textit{Adv. Theor. Math. Phys.} \textbf{2} (1999),
1249--1286, \eprint{hep-th/9808014}.
%
\bibitem{AGR}  Agricola I.,  Friedrich T.,
Killing spinors in supergravity with 4-fluxes, \textit{Class. Quantum
Grav.} \textbf{20} (2003), 4707--4717, \eprint{math/0307360}.
%
\bibitem{ACG} Alekseevsky  D.V., Cortes  V.,  Galaev A.S.,
Leistner T., Cones over pseudo-Riemannian manifolds and their
holonomy, \textit{J. Reine Angew. Math.} \textbf{635} (2009), 23--69,
\eprint{0707.3063}.
%
\bibitem{BAL}  Ballmann W., Lectures on K\"{a}hler Manifolds,
\textit{ESI Lectures in Mathematics and Physics}, Z\"{u}rich,
2006.
%
\bibitem{CB}
B\"ar C.,
Real Killing spinors and holonomy,
\textit{Comm. Math. Phys.} \textbf{154} (1993), 509--521.
%
\bibitem{BES}  Besse A., Einstein manifolds, Springer-Verlag,
New York, 1987.
%
\bibitem{BLR}  Blair D.E., Contact manifolds in Riemannian
geometry, \textit{Lecture Notes in Mathematics}, Vol.~509,
Springer-Verlag,
Berlin-New York, 1976.
%
\bibitem{BLZ}Bla\v{z}i\'{c}  N., Para-quaternionic projective
spaces and pseudo Riemannian geometry, \textit{Publ. Inst. Math.}
\textbf{60} (1996), 101--107.
%
\bibitem{BG}  Boyer C.,   Galicki K., 3-Sasakian manifolds,
Surveys in Differential Geometry: Essays on Einstein Manifolds,
\textit{Surv. Differ. Geom.}, Vol.~VI, 123--184, Int. Press, Boston,
MA, 1999, \eprint{hep-th/9810250}.
%
\bibitem{BG2} Boyer C.,   Galicki K., Sasakian geometry,
holonomy, and supersymmetry, Handbook of pseudo-Riemannian geometry
and supersymmetry, 39--83, \textit{IRMA Lect. Math. Theor. Phys.} 16,
Eur. Math. Soc., Z\"{u}rich, 2010, \eprint{math/0703231}.
%
\bibitem{BGM}  Boyer C.P.,  Galicki K.,  Mann B.M.,
The geometry and topology of 3-Sasakian manifolds,
\textit{J. Reine Angew. Math.}  \textbf{455} (1994), 183--220.
%
\bibitem{CP} Caldarella  A., Pastore  A.M., Mixed 3-Sasakian
structures and curvature, \textit{Ann. Polon. Math.} \textbf{96} (2009),
107--125,  \eprint{0803.1953}.
%
\bibitem{CKK}  Cariglia M., Krtou\v s P., Kubiz\v n\'{a}k D.,
Dirac equation in Kerr-NUT-(A)dS spacetimes: Intrinsic characterization
of separability in all dimensions, \eprint{1104.4123}.
%
\bibitem{BC}
Carter B.,
Killing tensor quantum numbers and conserved currents in curved spaces,
\textit{Phys. Rev. D} \textbf{16} (1977), 3395--3414.
%
\bibitem{CML} Carter B., McLenagham R.G.,
Generalized total angular momentum operator for Dirac equation in
curved space-time, \textit{Phys. Rev. D} \textbf{19} (1979),
1093--1097.
%
\bibitem{CLP}
Chen W.,  L\" u H.,  Pope C.N.,
General Kerr-NUT-AdS metrics in all dimensions,
\textit{Class. Quantum Grav.}  \textbf{23} (2006), 5323--5340,
\eprint{hep-th/0604125}.
%
\bibitem{CMMS}
Cort\'{e}s V., Mayer C., Mohaupt T.,
Saueressig F., Special geometry of euclidean supersymmetry II.
Hypermultiplets and the c-map, \textit{JHEP} \textbf{0506} (2005),
025, \eprint{hep-th/0503094}.
%
\bibitem{CLPP}
Cveti\v c M., L\"{u} H., Page D.N., Pope C.N.,
New Einstein-Sasaki and Einstein spaces from Kerr-de Sitter,
\textit{JHEP} \textbf{0907} (2005),
082, \eprint{hep-th/0505223}.
%
\bibitem{DW}
Dunajski M., West S., Anti-self-dual conformal
structures in neutral signature, \eprint{math.DG/0610280}.
%
\bibitem{ER}
Emparan R.,  Reall H.S.,
Black holes in higher dimensions,
\textit{Living Rev. Rel.} \textbf{11} (2008), 6, \eprint{0801.3471}.
%
\bibitem{GRM}  Garc\'{\i}a-R\'{\i}o E.,  Matsushita Y.,
 V\'{a}zquez-Lorenzo R.,
Paraquaternionic K\"{a}hler manifolds, \textit{Rocky Mt. J.
Math.} \textbf{31} (2001), 237--260.
%
\bibitem{GMSW1}
J. P. Gauntlett, D. Martelli, J. Sparks, D. Waldram,
Supersymmetric $AdS_5$ solutions of $M$-theory, 
\textit{Class. Quant. Grav.}  \textbf{21} (2004), 4335--4366,
\eprint{hep-th/0402153}.
%
\bibitem{GMSW2}
J. P. Gauntlett, D. Martelli, J. Sparks, D. Waldram,
Sasaki-Einstein metrics on $S^2 \times S^3$, 
\textit{Adv. Theor. Math. Phys.}  \textbf{8} (2004), 711--734,
\eprint{hep-th/0403002}.
%
\bibitem{GIB}  Gibbons G.W., Rychenkova P.,
Cones, tri-Sasakian structures and superconformal invariance,
\textit{Phys. Lett. B} \textbf{443} (1998), 138--142,
\eprint{hep-th/9809158}.
%
\bibitem{GHP}  Gibbons G.W.,  Hartnoll S.A.,  Pope C.N.,
Bohm and Einstein-Sasaki metrics, black holes and
cosmological event horizons,
\textit{Phys. Rev. D} \textbf{67} (2003), 084024  (23 pages),
\eprint{hep-th/0208031}.
%
\bibitem{HKWY1} 
Houri T., Kubiz\v n\'{a}k D., Warnick C., Yasui Y.,
Symmetries of the Dirac operator with skew-symmetric torsion,
\textit{Class. Quantum Grav.} \textbf{27} (2010), 185019,
\eprint{1002.3616}.
%
\bibitem{HKWY2} 
Houri T., Kubiz\v n\'{a}k D., Warnick C., Yasui Y.,
Generalized hidden symmetries and the Kerr-Sen black hole,
\textit{JHEP} \textbf{1007} (2010), 055,
\eprint{1004.1032}.
%
\bibitem{H1}
Hull C.M., Actions for (2,1) Sigma Models and
Strings, \textit{Nucl. Phys. B} \textbf{509} (1998), 252--272,
\eprint{hep-th/9702067}.
%
\bibitem{H2}
Hull C.M., Duality and the signature of space-time,
\textit{JHEP} \textbf{9811} (1998), 017,
\eprint{hep-th/0907127}.
%
\bibitem{IMV}  Ianu\c{s} S.,  Mazzocco R.,  V\^{\i}lcu G.E.,
Real lightlike hypersurfaces of paraquaternionic K\"{a}hler manifolds,
\textit{Mediterr. J. Math.} \textbf{3} (2006), 581--592.
%
\bibitem{IVV}  Ianu\c{s} S.,  Visinescu M.,  V\^{\i}lcu G.E.,
Conformal Killing-Yano tensors on manifolds with mixed
3-structures, \textit{SIGMA}
\textbf{5} (2009), Paper 022 (12 pages), \eprint{0902.3968}.
%
\bibitem{IV}  Ianu\c{s} S.,  V\^{\i}lcu G.E., Some constructions
of almost para-hyperhermitian structures on manifolds and tangent
bundles,
\textit{Int. J. Geom. Methods Mod. Phys.} \textbf{5} (2008), 893--903,
\eprint{0707.3360}.
%
\bibitem{IV2} Ianu\c{s} S.,  V\^{\i}lcu G.E.,
Paraquaternionic manifolds and mixed $3$-structures, in
\textit{Differential Geometry: Proceedings of the VIII International
Colloquium} (July 7-11, 2008, Santiago de Compostela, Spain),
Eds.Jes\'{u}s Alvarez L\'{o}pez and Eduardo Garc\'{\i}a-R\'{\i}o,
World Scientific Publishing Company, 2009, 276--285.
%
\bibitem{KN}  Kobayashi S.,  Nomizu K.,  Foundations of
Differential Geometry, 2 volumes, Interscience PUN., New York 1963,
1969.
%
\bibitem{DK}  Kubiz\v n\'{a}k D., On the supersymmetric limit
of Kerr-NUT-AdS metrics, \textit{Phys. Lett. B}  \textbf{675} (2009),
110--115, \eprint{0902.1999}.
%
\bibitem{KF}  Kubiz\v n\'{a}k D., Frolov V.P.,
Stationary strings and branes in the higher-dimensional Kerr-NUT-(A)dS
spacetimes, \textit{JHEP}  \textbf{0802} (2008), 007, \eprint{0711.2300}.
%
\bibitem{KUO} Kuo Y., On almost contact 3-structure,
\textit{Tohoku Math. J.} \textbf{22} (1970), 325--332.
%
\bibitem{KO}  Konishi M., On manifolds with Sasakian 3-structure over
quaternion Kaehler manifolds,
\textit{K\={o}dai Math. Semin. Rep.} \textbf{26} (1975), 194--200.
%
\bibitem{KKPV}
Krtou\v s P., Kubiz\v n\'{a}k D., Page D.N., Varsudevan M.,
Constants of geodesic motion in higher-dimensional black hole
spacetimes, \textit{Phys. Rev. D} \textbf{76} (2007), 084034,
\eprint{0707.0001}.
%
\bibitem{MS}
Martelli D.,  Sparks, J.,
Toric Sasaki-Einstein metrics on $S^2 \times S^3$,
\textit{Phys. Lett. B} \textbf{621} (2005), 208--212,
\eprint{hep-th/0505027}.
%
\bibitem{YO}
 Ohnita Y., Stability and rigidity of special Lagrangian cones
over certain minimal Legendrian orbits, \textit{Osaka J. Math.}
\textbf{44} (2007), 305--334.
%
\bibitem{OV}
Ooguri H., Vafa C., Geometry of N=2 strings,
\textit {Nucl. Phys. B} \textbf{361} (1991), 469--518.
%
\bibitem{OY1}  Oota T., Yasui Y.,
Separability of Dirac equation in higher dimensional Kerr-NUT-deSitter
spacetime, \textit{Phys. Lett. B} \textbf{659} (2008), 688--693,
\eprint{0711.0078}.
%
\bibitem{OY2}  Oota T., Yasui Y.,
Separability of Dirac gravitational perturbation in generalized
Kerr-NUT-deSitter spacetime, \textit{Int. J. Mod. Phys. A} \textbf{25}
(2008), 3055--3094, \eprint{0812.1623}.
%
\bibitem{PKVK}
Page D.N., Kubiz\v n\'{a}k D., Varsudevan M., Krtou\v s P.,
Complete integrability of geodesic motion in general Kerr-NUT-AdS
spacetimes, \textit{Phys. Rev. Lett.} \textbf{98} (2007), 061102,
\eprint{hep-th/0611083}.
%
\bibitem{SAS}  Sasaki S., On differentiable manifolds with
certain structures which are
closely related to almost contact structure I,
\textit{Tohoku Math. J.} \textbf{12} (1960), 459--476.
%
\bibitem{Sat}  Sato I., On a structure similar to the
almost contact structure, \textit{Tensor, New Ser.} \textbf{30} (1976),
219--224.
%
\bibitem{US} Semmelmann  U., Conformal Killing forms on Riemannian
manifolds, \textit{Math. Z.} \textbf{245} (2003), 503--527,
\eprint{math/0206117}.
%
\bibitem{SK}
Sergyeyev A., Krtou\v s P.,
Complete set of commuting symmetry operators for the Klein-Gordon
equation in generalized higher-dimensional Kerr-NUT-(A)dS spacetimes,
\textit{Phys. Rev. D} \textbf{77} (2008), 044033, \eprint{0711.4623}.
%
\bibitem{SS}  Sharfuddin A., Hasan Shahid  M., Hypersurfaces of
almost quaternion manifolds, \textit{Soochow J. Math.}
\textbf{20} (1994), 297--308.
%
\bibitem{SP}  Sparks J., Sasaki-Einstein Manifolds, \textit{Surv. Diff.
Geom.} \textbf{16} (2011), 265--324, \eprint{1004.2461}.
%
\bibitem{MV}
Visinescu M.,
Killing Forms on the Five-Dimensional Einstein-Sasaki Y(p,q) 
Spaces,
\textit{Mod. Phys. Lett. A} \textbf{27} (2012). 1250217 (8 pages),
\eprint{1207.2581}.
%
\bibitem{VV}   V\^{\i}lcu G.E.,  Voicu R.C., Curvature properties
of pseudo-sphere bundles over paraquaternionic manifolds,
\textit{ Int. J. Geom. Methods Mod. Phys.} \textbf{9} (2012), 1250024
(23 pages), \eprint{1107.5260}.
%
\bibitem{YAN} Yano  K., Some remarks on tensor fields and
curvature, \textit{Ann. of Math.} \textbf{55} (1952), 328--347.
%
\end{thebibliography}
\end{document}